\documentclass[sigconf]{acmart}
\usepackage{booktabs} 

\usepackage{subfig}
\usepackage{graphicx}
\usepackage{latexsym}
\usepackage{nidanfloat}
\usepackage{url}
\usepackage{amsmath}
\usepackage{amsfonts}
\usepackage{color}
\usepackage{clrs+}
\usepackage{flushend}

\newcommand{\procdecl}[1]   {\proc{#1}\vrule width0pt height0pt depth 7pt \relax}
\newcommand{\lilabel}[1]        {\label{li:#1}}
\newcommand{\liref}[1]      {line~\ref{li:#1}}

\newcommand{\erdosrenyi}{Erd\H os-R\'{e}nyi}

\newif\ifyusuke
\yusukefalse

\newcommand{\flop}{\mathrm{flop}}
\newcommand{\dnnz}{\textit{nnz}}

\def\fig_dir{.}

\emergencystretch=\maxdimen
\newlength\myindent
\setcopyright{none}
\setcopyright{rightsretained}

\renewcommand\footnotetextcopyrightpermission[1]{} 
\begin{document}
\title{High-performance sparse matrix-matrix products \\ on Intel KNL and multicore architectures}

\author{Yusuke Nagasaka}
\affiliation{%
  \institution{Tokyo Institute of Technology}
  \city{Tokyo}
  \country{Japan} 
}
\email{nagasaka.y.aa@m.titech.ac.jp}

\author{Satoshi Matsuoka}
\affiliation{%
  \institution{RIKEN Center for Computational Science}
  \city{Kobe}
  \country{Japan} 
}
\additionalaffiliation{
  \institution{Tokyo Institute of Technology}
  \department{Department of Mathematical and Computing Sciences}
  \city{Tokyo}
  \country{Japan}
}
\email{matsu@acm.org}

\author{Ariful Azad}
\affiliation{%
  \institution{Lawrence Berkeley National Laboratory}
  \city{Berkeley}
  \state{California}
  \country{USA}
}
\email{azad@lbl.gov}

\author{Ayd\i n Bulu\c{c}}
\affiliation{%
  \institution{Lawrence Berkeley National Laboratory}
  \city{Berkeley}
  \state{California}
  \country{USA}
}
\email{abuluc@lbl.gov}

\begin{abstract}
Sparse matrix-matrix multiplication (SpGEMM) is a computational primitive that is widely used in areas ranging from traditional numerical applications to recent big data analysis and machine learning. 
Although many SpGEMM algorithms have been proposed, hardware specific optimizations for multi- and many-core processors are lacking and a detailed analysis of their performance under various use cases and matrices is not available.
We firstly identify and mitigate multiple bottlenecks with memory management and thread scheduling on Intel Xeon Phi (Knights Landing or KNL). 
Specifically targeting multi- and many-core processors, 
we develop a hash-table-based algorithm and optimize a heap-based shared-memory SpGEMM algorithm. 
We examine their performance together with other publicly available codes.
Different from the literature, our evaluation also includes use cases that are representative of real graph algorithms, 
such as multi-source breadth-first search or triangle counting. 
Our hash-table and heap-based algorithms are showing significant speedups from libraries in the majority of the cases while different algorithms dominate the other scenarios with different matrix size, sparsity, compression factor and operation type.
We wrap up in-depth evaluation results and make a recipe to give the best SpGEMM algorithm for target scenario.
A critical finding is that hash-table-based SpGEMM gets a significant performance boost if the nonzeros are not required to be sorted within each row of the output matrix.
\end{abstract}

%
%


\maketitle

\section{Introduction}
Multiplication of two sparse matrices (SpGEMM) is a recurrent kernel in many algorithms in machine learning, data analysis, and graph analysis.
For example, bulk of the computation in multi-source breadth-first search~\cite{gapdt}, betweenness centrality~\cite{combblas}, Markov clustering~\cite{hipmcl}, label propagation~\cite{raghavan2007near}, peer pressure clustering~\cite{shahthesis}, clustering coefficients~\cite{triangles}, 
high-dimensional similarity search~\cite{agrawal2016exploiting}, and topological similarity search~\cite{he2010parallel} can be expressed as SpGEMM.
Similarly, numerical applications such as scientific simulations also use SpGEMM as a subroutine. 
Typical examples include the Algebraic Multigrid (AMG) method for solving sparse system of linear equations~\cite{ballard2016reducing}, volumetric
mesh processing~\cite{mueller2017ternary}, and linear-scaling electronic structure calculations~\cite{bock2016solvers}.

The extensive use of SpGEMM in data-intensive applications has led to the development of several sequential and parallel algorithms.
Most of these algorithms are based on Gustavson's row-wise SpGEMM algorithm~\cite{gustavson1978two} where a row of the output matrix is constructed by accumulating a subset of rows of the second input matrix (see Figure~\ref{algo:code_spgemm} for details). 
It is the accumulation (also called merging) technique that often distinguishes major classes of SpGEMM algorithms from one another.
Popular data structures for accumulating rows or columns of the output matrix include heap~\cite{azad2016exploiting}, hash~\cite{nagasaka2017high}, and sparse accumulator (SPA)~\cite{gilbert1992sparse}. 

Recently, researchers have developed parallel heap-, hash-, and SPA-based SpGEMM algorithms for shared-memory platforms~\cite{sulatycke1998caching, matam2012hipc, dalton2015toms, anh2016ics, gremse2015siam}. 
These algorithms are also packaged in publicly-available software that can tremendously benefit many scientific applications. 
However, when using an SpGEMM algorithm and implementation for a scientific problem, one needs answers to the following questions: (a) what is the best algorithm/implementation for a \emph{problem} at hand? (b) what is the best algorithm/implementation for the \emph{architecture} to be used in solving the problem? 
These practically important questions remain mostly unanswered for many scientific applications running on highly-threaded architectures.
This paper answers both questions in the context of existing SpGEMM algorithms. 
That means our focus is not to develop new parallel algorithms, but to characterize, optimize and evaluate existing algorithms for real-world applications on modern multicore and manycore architectures. This paper addresses the following gaps in the literature.

First, previous algorithmic work did not pay close attention to architecture-specific optimizations that have big impacts on the performance of SpGEMM.
We fill this gap by characterizing the performance of SpGEMM on shared-memory platforms and identifying bottlenecks in memory allocation and deallocation as well as overheads in thread scheduling. 
We propose solutions to mitigate those bottlenecks. Using microbenchmarks that model SpGEMM access patterns, we also uncover reasons behind the non-uniform performance boost provided by the MCDRAM on KNL. 
These optimizations resulted in efficient heap-based and hash-table-based SpGEMM algorithms that outperform state-of-the-art SpGEMM libraries including Intel MKL and Kokkos-Kernels~\cite{deveci2017performance} for many practical problems.

Second, previous work has narrowly focused on one or two real world application scenarios such as squaring a matrix and studying SpGEMM in the context of AMG solvers~\cite{patwary2015parallel, deveci2017performance}. 
Different from the literature, our evaluation also includes use cases that are representative of real graph algorithms, such as the multiplication of a square matrix with a tall skinny one that represent multi-source breadth-first search and the multiplication of triangular matrices that is used in triangle counting. 
While in the majority of the cases the hash-table-based SpGEMM algorithm is dominant, we also find that different algorithms dominate depending on matrix size, sparsity, compression factor, and operation type. 
This in-depth analysis exposes many interesting features of algorithms, applications, and multithreaded platforms. 

Third, while many SpGEMM algorithms keep nonzeros sorted within each row (or column) in increasing column (or row) identifiers, this is not universally necessary for subsequent sparse matrix operations. For example, CSparse~\cite{davis2006direct, davisprivate} assumes none of the input matrices are sorted. Clearly, if an algorithm accepts its inputs only in sorted format, then it must also emit sorted output for fairness. This is the case with the heap-based algorithms. However, hash based algorithm do not need their inputs sorted. In this case we see a significant performance benefit due to skipping the sorting of the output as well. 

Based on these architecture- and application-centric optimizations and evaluations, we make a recipe for selecting the best-performing algorithm for a specific application scenario. 
This recipe is based on empirical results, but also supported by theoretical performance estimations.
Therefore, this paper brings various SpGEMM algorithms and libraries together, analyzes them based on algorithm, application, and architecture related features and provides exhaustive guidelines for SpGEMM-dependent applications.

\section{Background and Related Work}
 Let $A, B$ be input matrices, and SpGEMM computes a matrix $C$ such that $C = AB$ where $A\in\mathbb{R}^{m \times n}$, $B\in\mathbb{R}^{n \times k}$, $C\in\mathbb{R}^{m \times k}$. The input and output matrices are sparse and they are stored in a sparse format. The number of nonzeros in matrix $A$ is denoted with $\dnnz(A)$. 
 Figure~\ref{algo:code_spgemm} shows the skeleton of the most commonly implemented SpGEMM algorithm, which is due to Gustavson~\cite{gustavson1978two}. When the matrices are stored using the Compressed Sparse Rows (CSR) format, this SpGEMM algorithm proceeds row-by-row on matrix $A$ (and hence on the output matrix $C$). Let $a_{ij}$ be the element in i-th row and j-th column of matrix $A$ and $a_{i*}$ be the i-th row of matrix $A$. The row of matrix $B$ corresponding to each non-zero element of matrix $A$ is read, and each non-zero element of output matrix $C$ is calculated.

\begin{figure}[!t]
  \begin{center}
  \begin{codebox}
      \Procname{$\procdecl{RowWise\_SpGEMM}(C,A,B)$}
\li \Comment set matrix C to $\emptyset$
\li \For $a_{i*}$ in matrix $A$ \InParallel
\li \Do \For $a_{ik}$ in row $a_{i*}$
\li \Do \For $b_{kj}$ in row $b_{k*}$
\li \Do $value \gets a_{ik}b_{kj}$
\li \If $c_{ij} \not\in c_{i*}$
\li \Then insert ($\id{c_{ij} \gets value}$)
\li \Else $\id{c_{ij} \gets c_{ij} + value}$ 
\End
 \EndDo
\end{codebox}
      \vspace{5pt}
  \end{center}
\caption{Pseudo code of Gustavson's Row-wise SpGEMM algorithm. The \InParallel keyword does not exist in the original algorithm but is used here to illustrate the common parallelization pattern of this algorithm used by all known implementations.}
      \vspace{-5pt}
\label{algo:code_spgemm}
\end{figure}

SpGEMM computation has three critical issues unlike dense matrix multiplication. First issue is indirect memory access to matrix data. As seen in Figure~\ref{algo:code_spgemm}, the memory access to matrix $B$ depends on the non-zero elements of matrix $A$. Therefore, the memory access to each non-zero element of matrix $B$ is indirect, and the cache miss may occur frequently. Secondly, the pattern and the number of non-zero elements of output matrix are not known beforehand. For this reason, the memory allocation of output matrix becomes hard, and we need to select from two strategies. One is a two-phase method, which counts the number of non-zero elements of output matrix first (symbolic phase), and then allocates memory and computes output matrix (numeric phase). The other is that we allocate large enough memory space for output matrix and compute. The former requires more computation cost, and the latter uses much more memory space. Finally, third issue is about combining the intermediate products ($value$ in Fig.~\ref{algo:code_spgemm}) to non-zero elements of output matrix.
Since the output matrix is also sparse, it is hard to efficiently accumulate intermediate products into non-zero elements. This procedure is a performance bottleneck of SpGEMM computation, and it is important to devise and select better accumulator for SpGEMM.

Since each row of $C$ can be constructed independently of each other, Gustavson's algorithm is conceptually highly parallel. 
For accumulation, Gustavson's algorithm uses a dense vector and a list of indices that hold the nonzero entries in the current active row. This particular set of data structures used in accumulation are later formalized by Gilbert et al.\, under the name of sparse accumulator (SPA)~\cite{gilbert1992sparse}. Consequently, a naive parallelization of Gustavson's algorithm requires temporary storage of $O(n\,t)$ where $t$ is the number of threads.  
For matrices with large dimensions, a SPA-based algorithm can still achieve good performance by ``blocking" SPA in order to decrease cache miss rates.
Patwary et al.\ \cite{patwary2015parallel} achieved this by partitioning the data structure of $B$ by columns.

Sulatycke and Ghose~\cite{sulatycke1998caching} presented the first shared-memory parallel algorithm for the SpGEMM problem, to the best of our knowledge. Their parallel algorithm, dubbed {\em IKJ method} due to the order of the loops, has a double-nested loop over the rows and the column of the matrix $A$. Therefore, the IKJ method has work complexity $O(n^2 + \flop)$ where $\flop$ is the number of the non-trivial scalar multiplications (i.e. those multiplications where both operands are nonzero) required to compute the product. Consequently, the IKJ method is only competitive when $\flop \geq n^2$, which is rare for SpGEMM. 

Several GPU algorithms that are also based on the row-by-row formulation are presented~\cite{liu2014ipdps, nagasaka2017high}. These algorithms first bin the rows based on their density due to the peculiarities of the GPU architectures. Then, a poly-algorithm executes a different specialized kernel for each bin, depending on its density. Two recent algorithms that are implemented in both GPUs and CPUs also follow the same row-by-row pattern, only differing on how they perform the merging operation. ViennaCL~\cite{rupp2016viennacl} implementation, which was first described for GPUs~\cite{gremse2015siam}, iteratively merges sorted lists, similar to merge sort. KokkosKernels implementation~\cite{deveci2017performance}, which we also include in our evaluation, uses a multi-level hash map data structure.

The CSR format is composed of three arrays: row pointers array (\id{rpts}) of length $n+1$, column indices (\id{cols}) of length $\id{nnz}$, and values (\id{vals}) of length $\id{nnz}$. Array \id{rpts} indexes the beginning and end locations of nonzeros within each row such that the range $\id{cols}[\id{rpts}[i] \ldots \id{rpts}[i+1]-1]$ lists the column indices of row $i$. The CSR format does not specify whether this range should be sorted with increasing column indices; that decision has been left to the library implementation. As we will show in our results, there are significant performance benefits of operating on unsorted CSR format. Table~\ref{tab:summary} lists high-level properties of the codes we study in this paper. Since MKL code is proprietary, we do not know the accumulator.

\begin{table}[h]
  \centering
  \caption{Summary of SpGEMM codes studied in this paper}
    \begin{tabular}{| l | c | c | c |}
\hline
{\bf Algorithm} & {\bf Phases} & {\bf Accumulator} & {\bf \shortstack{Sortedness \\ (Input/Output)}}\\
\hline
MKL & 2 & - & Any/Select \\
MKL-inspector & 1 & - & Any/Unsorted \\
KokkosKernels & 2 & HashMap & Any/Unsorted \\
Heap & 1 & Heap & Sorted/Sorted \\
Hash/HashVector & 2 & Hash Table & Any/Select \\
\hline
    \end{tabular}
  \label{tab:summary}
\end{table}

\section{Microbenchmarks on Intel KNL}
Our experiments target Intel Xeon and Xeon Phi architectures.
We conducted some preliminary experiments to tune optimization parameters to expose the performance bottlenecks.
These microbenchmarks are especially valuable for designing an algorithm of SpGEMM for these architectures.
Details of evaluation environment are summarized in Table~\ref{tab:cori}.

\subsection{Scheduling Cost of OpenMP}
\begin{figure}[t]
 \begin{center}
  \includegraphics[width=\hsize]{\fig_dir/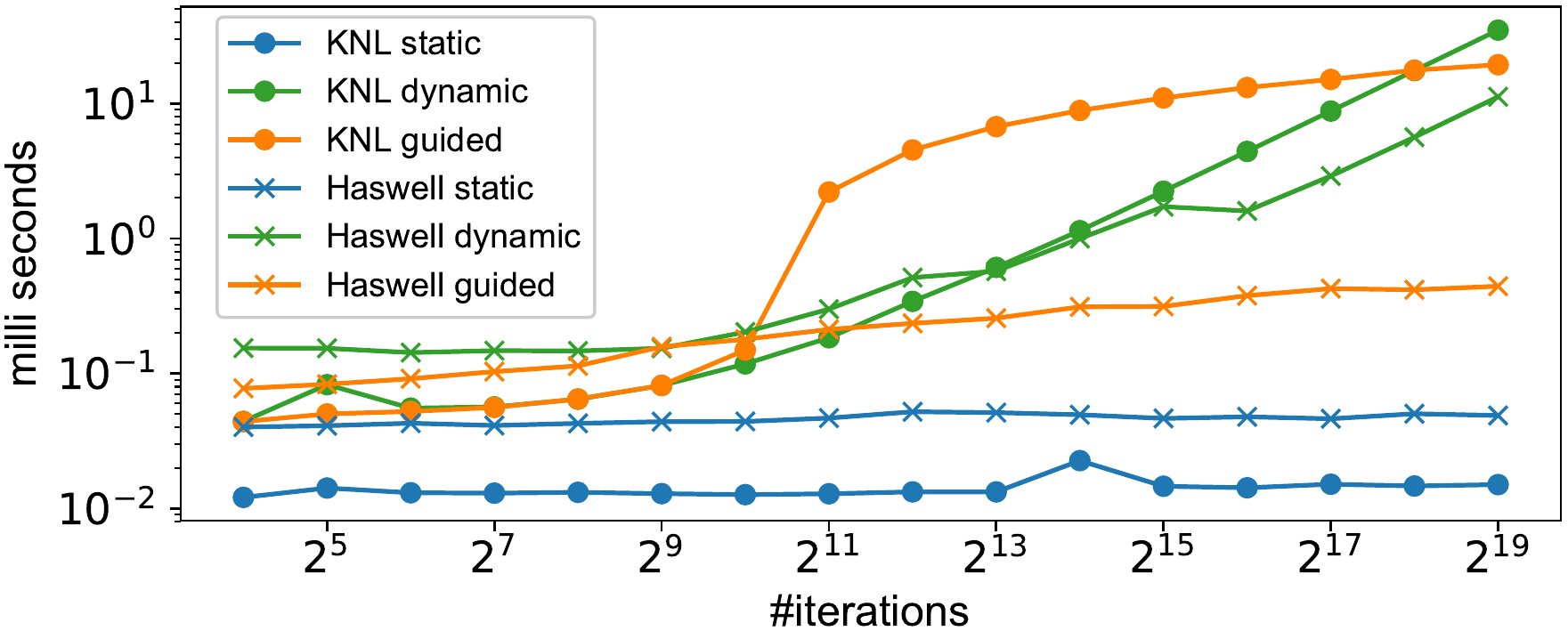}
  \caption{OpenMP Scheduling Cost on Haswell and KNL}
  \label{fig:omp_scheduling}
 \end{center}
\end{figure}

When parallelizing a loop, OpenMP provides three thread scheduling choices: {\it static}, {\it dynamic} and {\it guided}. 
Static scheduling divides loop iterations equally among threads, dynamic scheduling allocates iterations to threads dynamically, and 
guided scheduling starts with static scheduling with smaller iterations and switch to dynamic scheduling in the later part.
Here, 
we experimentally evaluate the scheduling cost of these three scheduling options on Haswell and KNL processors by running simple program, which only repeats loop iterations without doing anything in the loop. We measure the time during loop iterations and the result is shown in Figure~\ref{fig:omp_scheduling}.
In load-balanced situation with a large number of iterations, 
static scheduling takes very little scheduling overhead than dynamic scheduling on both Haswell and KNL, as expected.
The guided scheduling is also as expensive as dynamic scheduling, especially on the KNL processor.
Based on these evaluations, we opt to use static scheduling in our SpGEMM algorithms since the scheduling overhead of dynamic or guided scheduling becomes an bottleneck.

\subsection{Memory Allocation/Deallocation}
\begin{figure}[t]
 \begin{center}
  \includegraphics[width=\hsize]{\fig_dir/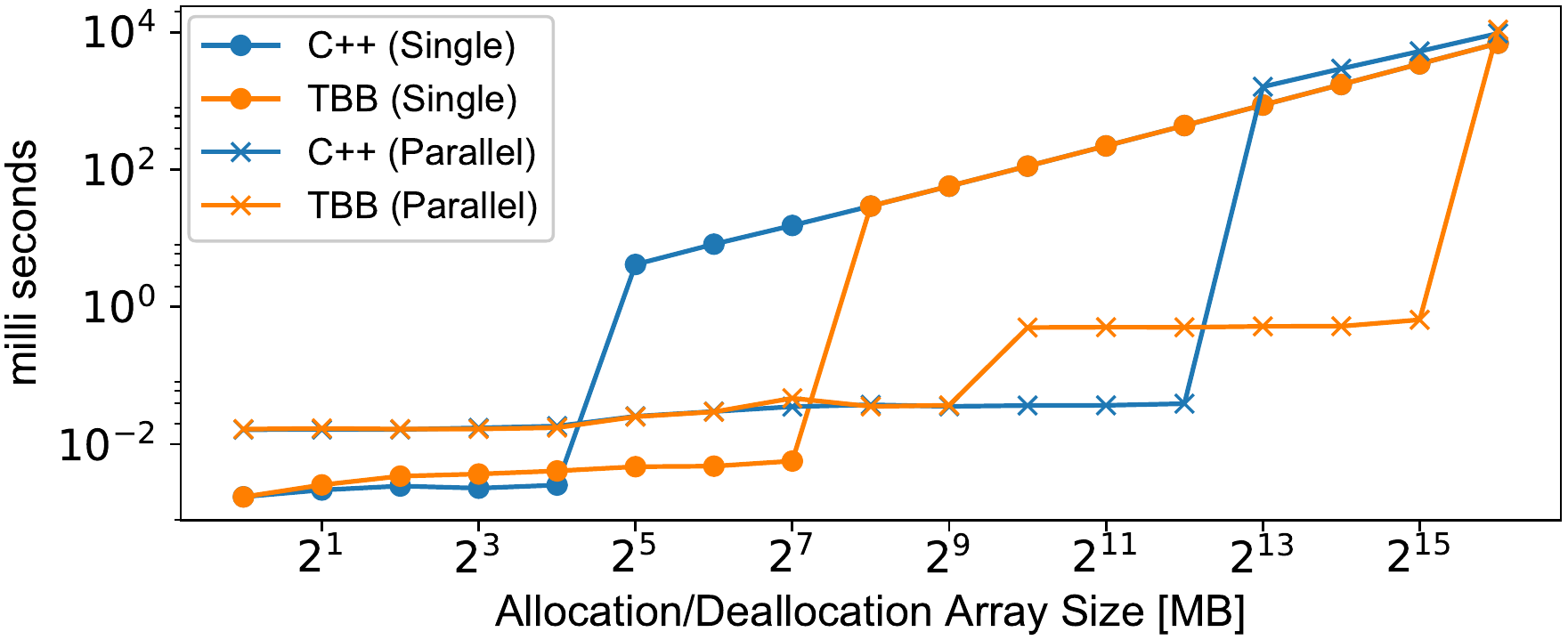}
  \caption{Cost of deallocation on KNL}
  \label{fig:knl_deallocation}
 \end{center}
\end{figure}

To find a suitable memory allocation/deallocation scheme on KNL, we performed a simple experiment: allocate a memory space, access elements on the allocated memory and then deallocate it.
To contrast this ``single'' memory management scheme, we considered a ``parallel'' approach where each thread independently allocates/deallocates equal portion of the total requested memory and accesses only its own allocated memory space.
The latter approach is described in Figure~\ref{algo:knl_parallel_allocation}.
We examined three ways to allocate or deallocate memory; new/delete of C++, aligned allocation (\_mm\_malloc/\_free), and scalable\_malloc/\_free provided by Intel TBB (Thread Building Block). 
Figure~\ref{fig:knl_deallocation} shows the results of ``single'' deallocation and ``parallel'' deallocation with 256 threads on KNL. 
Since aligned allocation showed nearly same performance as C++, we show only the results of C++ and TBB.
All ``single'' allocators have extremely high cost for large memory: over 100 milliseconds for the deallocation of 1GB memory space. The ``parallel'' deallocation for large memory chunks is much cheaper than ``single'' deallocation. The cost of ``parallel'' deallocation suddenly rises at 8GB (C++) or 64GB (TBB), where each thread allocates 32MB or 256MB, respectively. These thresholds match those of ``single'' deallocation.
On the other hand, the cost of ``parallel'' deallocation for small memory space becomes larger than ``single'' deallocation since ``parallel'' deallocation causes the overheads of OpenMP scheduling and synchronization.
From this result, we compute the amount of memory required by each thread and allocate this amount of thread-private memory in each thread independently in order to reduce deallocation cost in SpGEMM computation, which requires temporally memory allocation and deallocation.
In the following experiments in this paper, the TBB is used for both ``single'' and ``parallel'' memory allocation/deallocation to simply have performance gain.
\begin{figure}[t]
\begin{codebox}
\li $\id{eachN} \gets N/{\id{nthreads}}$
\li \proc{Allocate}($\id{a}, \id{nthreads})$
\li \For $\id{tid} \gets$ \To $\id{nthreads}$ \InParallel
\li \Do \proc{Allocate}($\id{a[tid]}, \id{eachN})$
\End
\li \Do \For $\id{i} \gets$ \To $\id{eachN}$
\li \Do $\id{a[tid][i] \gets i}$
\End
\End
\li \Do \proc{Deallocate}($\id{a[tid]},\id{eachN})$
\End
\li \proc{Deallocate}($\id{a[tnum]})$
\end{codebox}
\caption{Experiment of Parallel Memory Management. $a$ is an array of arrays. Total requested memory is $N$, and each thread independently requests equal size, $eachN$.}
\label{algo:knl_parallel_allocation}
\end{figure}

\subsection{DDR vs. MCDRAM}
Intel KNL implements MCDRAM, which can accelerate bandwidth-bound algorithms and applications. While MCDRAM provides high bandwidth, its memory latency is larger than that of DDR4.
In row-wise SpGEMM (Algorithm~\ref{algo:code_spgemm}), there are three main types of data accesses for the formation of each row of $C$. First, there is a unit-stride streaming access pattern arising from access of the row pointers of $A$ as well as the creation of the sparse output vector $c_{i*}$. Second, access to rows of $B$ follows a stanza-like memory access pattern where small blocks (stanzas) of consecutive elements are fetched from effectively random locations in memory. Finally, updates to the accumulator exhibit different access pattern depending on the type of the accumulator (a hash table, SPA, or heap). The streaming access to the input vector is usually the cheapest of the three and the accumulator access depends on the data structure used. Hence, stanza access pattern is the most canonical of the three and provides a decent proxy to study.

\begin{figure}[t]
 \begin{center}
  \includegraphics[width=\hsize]{\fig_dir/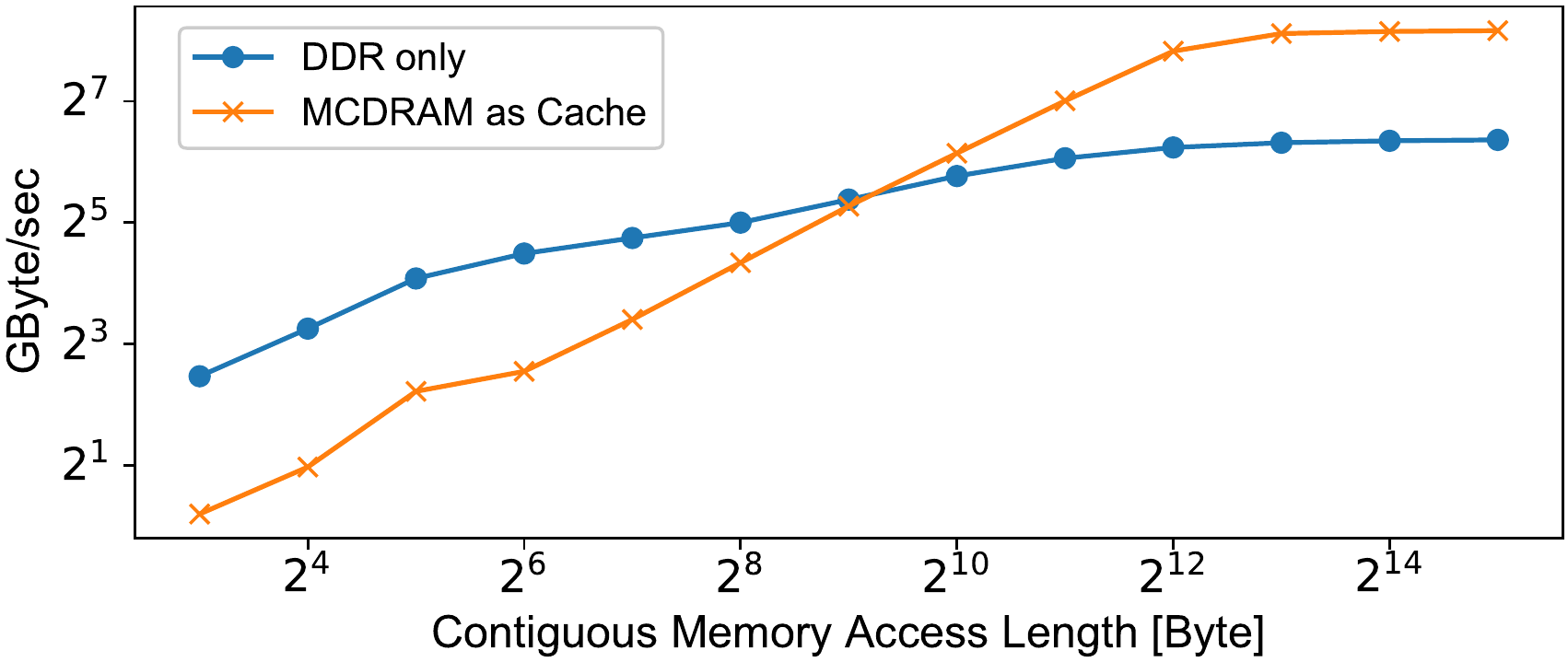}
  \caption{Benchmark result of random memory access with DDR only or MCDRAM as Cache}
  \label{fig:stanza}
 \end{center}
\end{figure}

To quantify the stanza bandwidth which we expect to be quite different than STREAM~\cite{McCalpin2007}, we used a custom microbenchmark that provides stanza-like memory access patterns (read or update) with spatial locality varying from 8 bytes (random access) to the size of the array (i.e. asymptotically the STREAM benchmark). 
Figure~\ref{fig:stanza} shows a comparison result between DDR only and use of MCDRAM as Cache with scaling the length of contiguous memory access. When the contiguous memory access is wider, both DDR only and MCDRAM as Cache achieve their peak bandwidth, and especially MCDRAM as Cache shows over $3.4\times$ superior bandwidth compared to DDR only. However, the use of MCDRAM as Cache is incompatible with fine-grained memory access. When the stanza length is small, there is little benefit of using MCDRAM. This benchmark hints that it would be hard to get the benefits of MCDRAM on very sparse matrices.

\section{Architecture Specific Optimization of SpGEMM}
Based on our findings of performance critical bottlenecks on Intel KNL, we design SpGEMM algorithms taking account in architecture specific issues.
First, we show light-weight  thread scheduling scheme with load-balancing for SpGEMM.
Next, we show the optimization schemes for hash table based SpGEMM, which is proposed for GPU~\cite{nagasaka2017high}, and heap based shared-memory SpGEMM algorithms~\cite{azad2016exploiting}. Additionally, we extend the Hash SpGEMM with utilizing vector registers of Intel Xeon or Xeon Phi.
Finally, we show which accumulator works well for target scenario from the theoretical point of view by estimating each accumulation cost.

\subsection{Light-weight Load-balancing Thread Scheduling Scheme}
To achieve good load-balance with static scheduling, the bundle of rows should be assigned to threads with equal computation complexity before symbolic or numeric phase.
Figure~\ref{algo:load_balance} shows how to assign rows to threads. First, we count $\flop$ of each row, then do prefix sum. Each thread can find the start point of rows by binary search. $\proc{lowbnd}(\id{vec}, \id{value})$ in \liref{lowbound<} finds the minimum $\id{id}$ such that $\id{vec[id]}$ is larger than or equal to \id{value}.
Each of these operations can be executed in parallel.
\begin{figure}[!t]
  \begin{center}
\begin{codebox}
    \Procname{$\procdecl{RowsToThreads}(\id{offset},A,B)$}
\li \Comment {1. Set FLOPS vector}
\li \For $\id{i} \gets 0$ \To $\id{m}$ \InParallel
\li \Do $\id{\flop [i]} \gets 0$
\li \For $\id{j} \gets \id{rpts_A[i]}$ \To $\id{rpts_A}[i+1]$
\li \Do $\id{rnz} \gets \id{rpts_B}[\id{cols_A}[j] + 1] - \id{rpt_B}[\id{cols_A}[j]]$
\li $\flop [i] \gets \flop [i] + \id{rnz}$
\End
\End
\li \Comment {2. Assign rows to thread}
\li $\id{\flop_{ps}} \gets \proc{ParallelPrefixSum}(\flop)$
\li $\id{sum_{\flop} \gets \flop_{ps}[m]}$
\li $\id{tnum} \gets \proc{omp\_get\_max\_threads}()$
\li $\id{ave_{\flop}} \gets \id{sum}_{\flop}/\id{tnum}$
\li $\id{offset[0]} \gets 0$
\li \For $\id{tid} \gets 1$ \To $\id{tnum}$ \InParallel
\li \Do 
 $\id{offset}[\id{tid}] \gets \proc{lowbnd}(\flop_{ps}, \id{ave_{\flop}} * \id{tid})$ \lilabel{lowbound<}
 \End
\li $\id{offset[tnum] \gets m}$
\end{codebox}
  \end{center}
\caption{Load-balanced Thread Assignment}
\label{algo:load_balance}
\end{figure}

\subsection{Symbolic and Numeric Phases}
We optimized two approaches of accumulation for KNL, one is hash-table based algorithm and the other is heap based algorithm. Furthermore, we add another version of Hash SpGEMM, where hash probing is vectorized with AVX-512 or AVX2 instructions.

\subsubsection{Hash SpGEMM}
\begin{figure}[t!]
 \begin{center}
  \includegraphics[width=\hsize]{\fig_dir/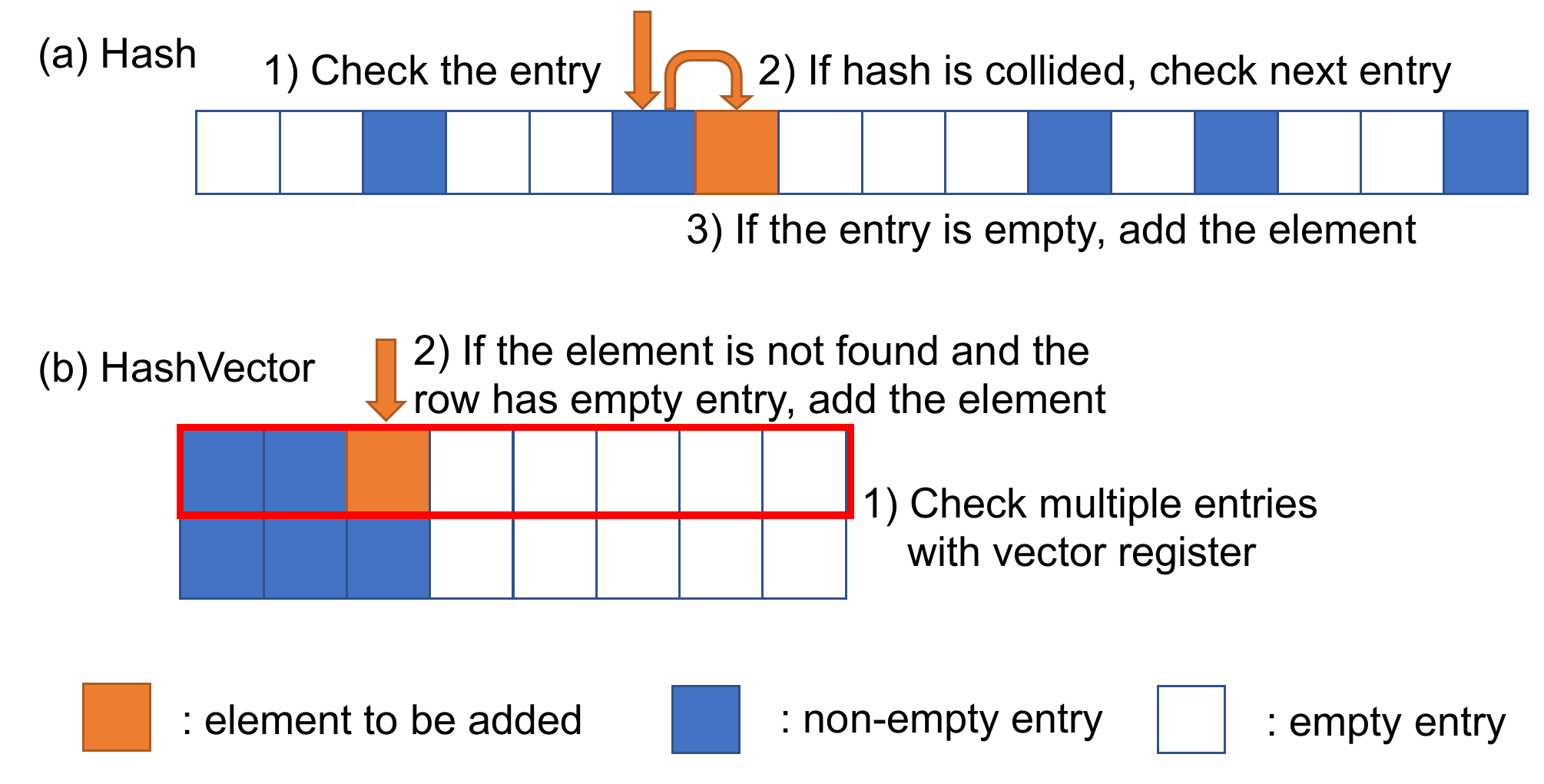}
  \caption{Hash Probing in Hash and HashVector SpGEMM}
  \label{fig:hash_algorithm}
 \end{center}
\end{figure}

We use hash table for accumulator in SpGEMM computation, based on GPU work~\cite{nagasaka2017high}. Figure~\ref{algo:hash_spgemm} shows the algorithm of Hash SpGEMM for multi- and many-core processors. 
The allocation/deallocation scheme of hash table for each thread is based on ``parallel'' approach like FIgure~\ref{algo:knl_parallel_allocation}.
We count a $\flop$ per row of output matrix. The upper limit of any thread's local hash table size is the maximum number of $\flop$ per row within the rows assigned to it. Each thread once allocates the hash table based on its own upper limit and reuses that hash table throughout the computation by reinitializing for each row. Next is about hashing algorithm we adopted. A column index is inserted into hash table as key. Since the column index is no less than 0, the hash table is initialized by storing $-1$.  The column index is multiplied by constant number and divided by hash table size to compute the remainder. In order to compute modulus operation efficiently, the hash table size is set as $2^n$ ($n$ is a integer). The hashing algorithm is based on linear probing. Figure~\ref{fig:hash_algorithm}-(a) shows an example of hash probing on 16 entries hash table. 

In symbolic phase, it is enough to insert keys to the hash table. In numeric phase, however, we need to store the resulting value data. 
Once the computation on the hash table finishes, the results are sorted by column indices in ascending order (if necessary), and stored to memory as output.
This Hash SpGEMM for multi/many-core processors differs from the GPU version as follows. While a row of output is computed by multiple threads in the GPU version to exploit massive number of threads on GPU, each row is processed by a single thread in the present algorithm. Hash SpGEMM on GPU requires some form of mutual exclusion since multiple threads access the same entry of the hash table concurrently. We were able to remove this overhead in our present Hash SpGEMM for multi/many-core processors.

\begin{figure}[!t]
  \begin{center}
\begin{codebox}
      \Procname{$\procdecl{Hash\_SpGEMM}(C,A,B)$}
\li $\proc{RowsToThreads}(\id{offset}, A,B)$
\li \proc{Allocate}($\id{table_{id}}, tnum)$
\li \proc{Allocate}($\id{table_{val}}, tnum)$
\li \Comment Determine hash table size for each thread
\li \For $\id{tid} \gets 0$ \To $\id{tnum}$
\li \Do $\id{size_t \gets 0}$
\li \For $\id{i \gets offset[tid]}$ \To $\id{offset[tid + 1]}$
\li \Do $\id{size_t} \gets \proc{max}(\id{size_t}, \id{\flop[i]})$
\End
\li \Comment Required maximum hash table size is $\id{N_{col}}$
\li $\id{size_t} \gets \proc{min}(\id{N_{col}}, \id{size_t})$
\li \Comment Return minimum $\id{2^n}$ so that $\id{2^n} > \id{size_t}$
\li $\id{size_t} \gets \proc{lowest\_p2}(\id{size_t})$
\li \proc{Allocate}($\id{table_{id}[tid], size_t}$)
\li \proc{Allocate}($\id{table_{val}[tid], size_t}$)
\End
\li \proc{Allocate}($\id{rpts_C, \id{N_{row}} + 1}$)
\li \proc{Symbolic}($\id{rpts_C}, A, B$)
\li \proc{Allocate}($\id{cols_C, rpts_C[\id{N_{row}}]}$)
\li \proc{Allocate}($\id{vals_C, rpts_C[\id{N_{row}}]}$)
\li \proc{Numeric}($\id{C, A, B}$)
\end{codebox}
      \vspace{5pt}
  \end{center}
\caption{Hash SpGEMM Pseudocode}
\label{algo:hash_spgemm}
\end{figure}

\subsubsection{HashVector SpGEMM}
Intel Xeon or Xeon Phi implements 256 and 512-bit wide vector register, respectively. This vector register reduces instruction counts and brings large benefit to algorithms and applications, which require contiguous memory access. However, sparse matrix computation has indirect memory access, and hence it is hard to utilize vector registers.
In this paper, we utilize vector register for hash probing in our Hash SpGEMM algorithm. The vectorization of hash probing is based on Ross~\cite{ross2007efficient}. Figure~\ref{fig:hash_algorithm}-(b) shows how HashVector algorithm works hash probing.
The size of hash table is 16, same as (a), and it is divided into chunks based on vector width.
A chunk consists of 8 entries on Haswell since a key (= column index) is represented as 32-bit in our evaluation.
In HashVector, the hash indicates the identifier of target chunk in hash table. In order to examine the keys in the chunk, we use comparison instruction with vector register.
If the entry with target key is found, the algorithm finishes the probing for the element in symbolic phase. In numeric phase, the target entry in chunk is identified by \_\_builtin\_ctz function, which counts trailing zeros, and the multiplied value is added to the value of the entry.
If the algorithm finds no entry with the key, the element is pushed to the hash table. In HashVector, new element is pushed into the table in order from the beginning. The entries in chunk are compared with the initial value of hash table, -1, by using vector register. The beginning of empty entries can be found by counting the number of bit flags of comparison result.
When the chunk is occupied with other keys, the next chunk is to be checked in accordance with linear probing.
Since Hash vector SpGEMM can reduce the number of probing caused by hash collision, it can achieve better performance compared to Hash SpGEMM. However, HashVector requires a few more instructions for each check. Thus, HashVector may degrade the performance when the collisions in Hash SpGEMM are rare.

\subsubsection{Heap SpGEMM}
In another variant of SpGEMM~\cite{azad2016exploiting}, 
we use a priority queue (heap) -- indexed by column indices -- to accumulate each row of $C$. 
To construct $c_{i*}$, a heap of size  $\dnnz(a_{i*})$ is allocated.
For every nonzero $a_{ik}$, the first nonzero entry in $b_{k*}$ along with its column index is inserted into the heap.
The algorithm iteratively extracts an entry with the minimum column index from the heap, accumulates it to $c_{i*}$, and inserts the next nonzero entry from the last  extracted row of $B$ into the heap.
When the heap becomes empty, the algorithm moves to construct the next row of $C$.

Heap SpGEMM can be more expensive than hash- and SPA- based algorithms because it requires logarithmic time to extract elements from the heap.
However, from the accumulator point of view, Heap SpGEMM is space efficient as it uses $O(\dnnz(a_{i*}))$ memory to accumulate $c_{i*}$ instead of $O(\flop(c_{i*}))$ and $O(n)$ memory used by hash- and SPA-based algorithms, respectively.

Our implementation of Heap SpGEMM adopts the one-phase method, which require larger memory usage for temporally keeping the output. In parallel Heap SpGEMM, because rows of $C$ are independently constructed by different threads, this temporary memory use for keeping output is thread-independent and we can adapt ``parallel'' approach for memory management. Thread-private heaps are also managed with ``parallel'' approach.
As with the Hash algorithm, Heap SpGEMM estimates $\flop$ via a symbolic step and uses it to balance computational load evenly among threads. 

\subsubsection{Performance Estimation for A Recipe}
\label{sec:perf_est}
To estimate how our algorithms would perform in practice, we show their computation costs.
As described, Heap SpGEMM requires logarithmic time to extract elements to the heap. The complexity of Heap SpGEMM is:
\begin{equation}
\label{eq:heap_est}
T_{heap} = \sum_{i=1}^{n} \left( \flop(c_{i*}) * \log \dnnz(a_{i*}) \right)
\end{equation}
On the other hand, Hash SpGEMM requires $O(1)$ cost to explore in its hash table if there is no hash collision. We introduce $c$, collision factor, which is an average number of probing to detect/insert the key. When $c=1$, no hash collision occurs during SpGEMM computation. In addition to this hash probing cost, Hash SpGEMM requires sorting, which takes $O(n \log n)$ time, if an application needs. The computational complexity of Hash SpGEMM is:
\begin{equation}
\label{eq:hash_est}
T_{hash} = \flop * c + \sum_{i=1}^{n} \left( \dnnz(c_{i*}) * \log \dnnz(c_{i*}) \right)
\end{equation}

From (\ref{eq:heap_est}) and (\ref{eq:hash_est}), Hash SpGEMM tends to achieve superior performance to Heap SpGEMM when $\dnnz(c_{i*})$ or $\flop(c_{i*}) / \dnnz(c_{i*})$ is large. Denser input matrices make output matrix denser too. Also, the multiplication of input matrices with regular non-zero patterns output a regular matrix, and in that case, $\flop(c_{i*}) / \dnnz(c_{i*})$ is large. Thus, we can guess that Hash becomes better choice when input matrices are dense or have regular structure.

\section{Experimental Results}
Different SpGEMM algorithms can dominate others depending on the aspect ratio (i.e.\, ratio of its dimensions), density, sparsity structure, and size (i.e. dimensions) of its inputs.
To provide a comprehensive and fair assessment, we evaluate SpGEMM codes under several different scenarios.
For the case where input and output matrices are sorted, we evaluate MKL, Heap and Hash/HashVector, and for the case where they are unsorted we evaluate MKL, MKL-inspector, KokkosKernels (with `kkmem' option) and Hash/HashVector.
Each performance number in the following part is the average of ten SpGEMM runs.

\subsection{Input Types}
We use two types of matrices for the evaluation. We generate synthetic matrix using matrix generator, and take matrices from SuiteSparse Matrix Collection (Formerly University of Florida Sparse Matrix Collection)~\cite{davis2011university}.
For the evaluation of unsorted output, the column indices of input matrices are randomly permuted. 
We use 26 sparse matrices used in~\cite{liu2014ipdps, triangles, deveci2017performance}. The matrices are listed in Table~\ref{tab:florida_mat}.

We use R-MAT~\cite{chakrabarti2004r}, the recursive matrix generator, to generate two different non-zero patterns of synthetic matrices represented as ER and G500. ER matrix represents \erdosrenyi\ random graphs, and G500 represents graphs with power-law degree distributions used for Graph500 benchmark. These matrices are generated with R-MAT seed parameters; $a=b=c=d=0.25$ for ER matrix and $a=0.57, b=c=0.19, d=0.05$ for G500 matrix.
A scale $n$ matrix represents $2^n$-by-$2^n$. The {\em edge factor} parameter for these generators is the average number of non-zero elements per row (or column) of the matrix. In other words, it is the ratio of \dnnz\ to n. 

\begin{table}[]
  \centering
  \caption{Matrix data used in our experiments (all numbers are in millions)}
    \begin{tabular}{| l | r | r | r | r |}
\hline
{\bf Matrix} & {\bf n} & {\bf $\dnnz(A)$} & {$\mathrm{\bf flop}$}{\bf ($A^2$)} & {\bf $\dnnz(A^2)$} \\
\hline
2cubes\_sphere & 0.101  & 1.65  & 27.45  & 8.97  \\
cage12 & 0.130  & 2.03  & 34.61  & 15.23  \\
cage15 & 5.155  & 99.20  & 2,078.63  & 929.02  \\
cant & 0.062  & 4.01  & 269.49  & 17.44  \\
conf5\_4-8x8-05 & 0.049  & 1.92  & 74.76  & 10.91  \\
consph & 0.083  & 6.01  & 463.85  & 26.54  \\
cop20k\_A & 0.121  & 2.62  & 79.88  & 18.71  \\
delaunay\_n24 & 16.777  & 100.66  & 633.91  & 347.32  \\
filter3D & 0.106  & 2.71  & 85.96  & 20.16  \\
hood & 0.221  & 10.77  & 562.03  & 34.24  \\
m133-b3 & 0.200  & 0.80  & 3.20  & 3.18  \\
mac\_econ\_fwd500 & 0.207  & 1.27  & 7.56  & 6.70  \\
majorbasis & 0.160  & 1.75  & 19.18  & 8.24  \\
mario002 & 0.390  & 2.10  & 12.83  & 6.45  \\
mc2depi & 0.526  & 2.10  & 8.39  & 5.25  \\
mono\_500Hz & 0.169  & 5.04  & 204.03  & 41.38  \\
offshore & 0.260  & 4.24  & 71.34  & 23.36  \\
patents\_main & 0.241  & 0.56  & 2.60  & 2.28  \\
pdb1HYS & 0.036  & 4.34  & 555.32  & 19.59  \\
poisson3Da & 0.014  & 0.35  & 11.77  & 2.96  \\
pwtk & 0.218  & 11.63  & 626.05  & 32.77  \\
rma10 & 0.047  & 2.37  & 156.48  & 7.90  \\
scircuit & 0.171  & 0.96  & 8.68  & 5.22  \\
shipsec1 & 0.141  & 7.81  & 450.64  & 24.09  \\
wb-edu & 9.846  & 57.16  & 1,559.58  & 630.08  \\
webbase-1M & 1.000  & 3.11  & 69.52  & 51.11  \\
\hline
    \end{tabular}
  \label{tab:florida_mat}
\end{table}

\subsection{Experimental Environment}
We evaluate the performance of SpGEMM on a single node of the Cori supercomputer at NERSC. Cori system consists of two partitions; one is Intel Xeon Haswell cluster (Haswell), and the other is Intel KNL cluster. We use nodes from both partitions of Cori. Details are summarized in Table~\ref{tab:cori}.

The Haswell and KNL processors provide hyperthreading with 2 or 4 threads for each core respectively. We set the number of threads as 68, 136, 204 or 272 for KNL, and 16, 32 or 64 for Haswell. For the evaluation of Kokkos on KNL, we set 256 threads instead of 272 threads since the execution fails on more than 256 threads. We show the result with the best thread count. 
For the evaluation on KNL, we set ``quadrant'' cluster mode, and mainly ``Cache'' memory mode. To select DDR4 or MCDRAM with ``Flat'' memory mode, we use ``numactl -p''. The thread affinity is set as ``KMP\_AFFINITY=`granularity=fine',scatter''.

\begin{table}[]
  \centering
  \caption{Overview of Evaluation Environment (Cori system)}
    \begin{tabular}{ l  l  l}
 & {\bf Haswell cluster} & {\bf KNL cluster} \\
\hline
{\bf CPU} & {\bf \shortstack[l]{Intel Xeon \\ Processor E5-2698 v3}} & {\bf \shortstack[l]{Intel Xeon Phi \\ Processor 7250}} \\
\hline
\#Sockets & 2 & 1 \\
\#Cores/socket & 16 & 68 \\
Clock & 2.3GHz & 1.4GHz \\
L1 cache & 32KB/core & 32KB/core \\
L2 cache & 256KB/core & 1MB/tile \\
L3 cache & 40MB per socket & - \\
\hline
{\bf Memory} & & \\
\hline
DDR4 & 128GB & 96GB \\
MCDRAM & - & 16GB \\
\hline
{\bf Software} & \\
\hline
OS & \multicolumn{2}{c}{SuSE Linux Enterprise Server} \\
Compiler & \multicolumn{2}{c}{Intel C++ Compiler (icpc) ver18.0.0} \\
Option & \multicolumn{2}{c}{-g -O3 -qopenmp} \\
\hline
    \end{tabular}
  \label{tab:cori}
\end{table}

\subsection{Preliminary Evaluation on KNL}
\subsubsection{Advantage of Performance Optimization on KNL for SpGEMM}
\begin{figure}[t]
 \begin{center}
    \includegraphics[width=\hsize]{\fig_dir/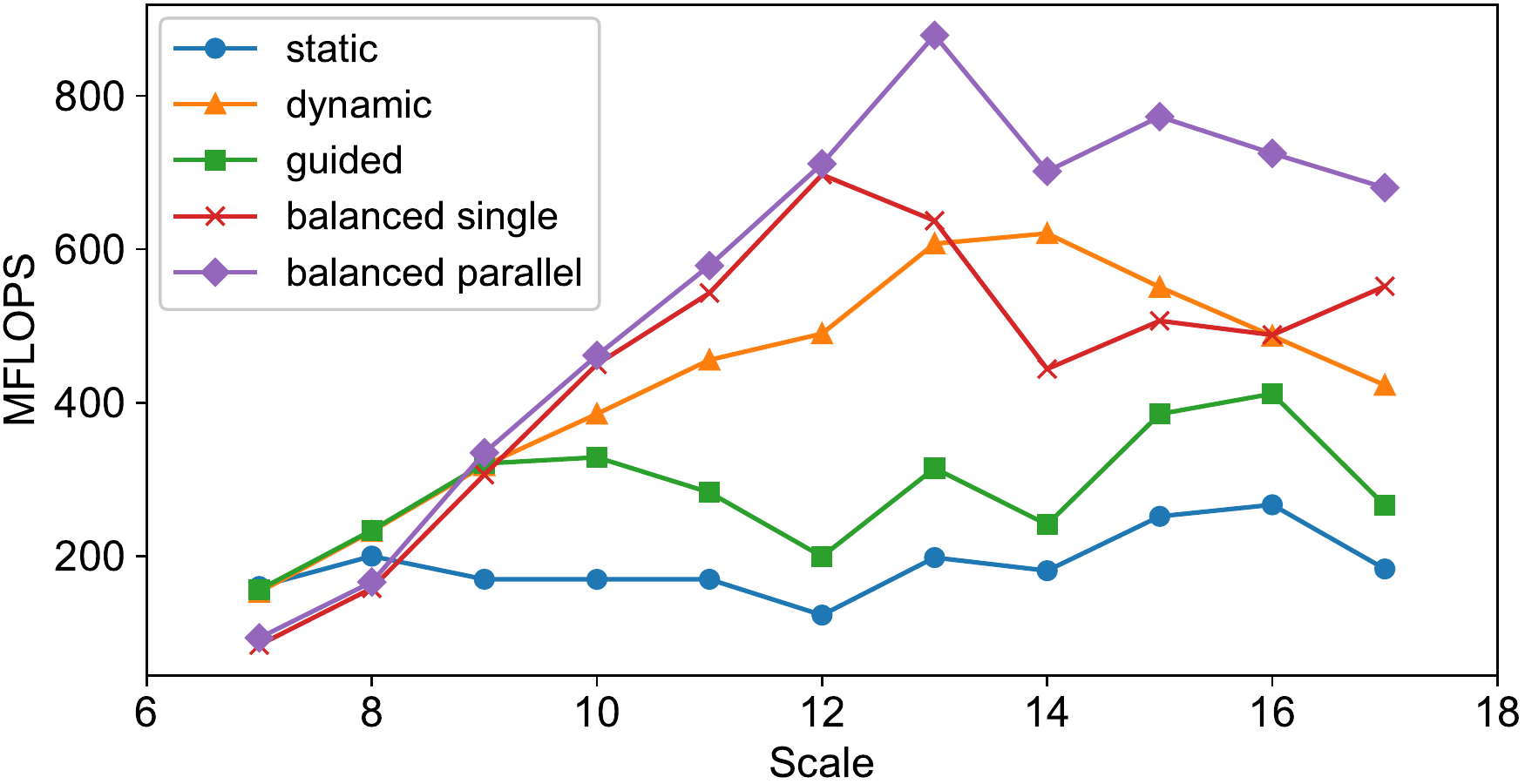}
  \caption{Performance of Heap SpGEMM scaling with size of G500 inputs on KNL with Cache mode}
  \label{fig:spgemm_breakdown}
 \end{center}
\end{figure}

We examined performance difference between OpenMP scheduling and ways to allocate memory. Figure~\ref{fig:spgemm_breakdown} shows the performance of Heap SpGEMM for squaring G500 matrices with edge factor 16. When simply parallelizing SpGEMM by row, we cannot achieve higher performance because of load imbalance with static scheduling or expensive scheduling overhead with dynamic/guided scheduling. On the other hand, our light-weight load-balancing thread scheduling scheme, `balanced', works well on SpGEMM.
For larger inputs, Heap SpGEMM temporally requires large memory use, whose deallocation causes performance degradation. Our ``parallel'' memory management scheme `balanced parallel' reduces the overhead of memory deallocation for temporal memory use, and keeps high performance on large size inputs.

\subsubsection{DDR vs MCDRAM}
\begin{figure}[t]
 \begin{center}
    \includegraphics[width=\hsize]{\fig_dir/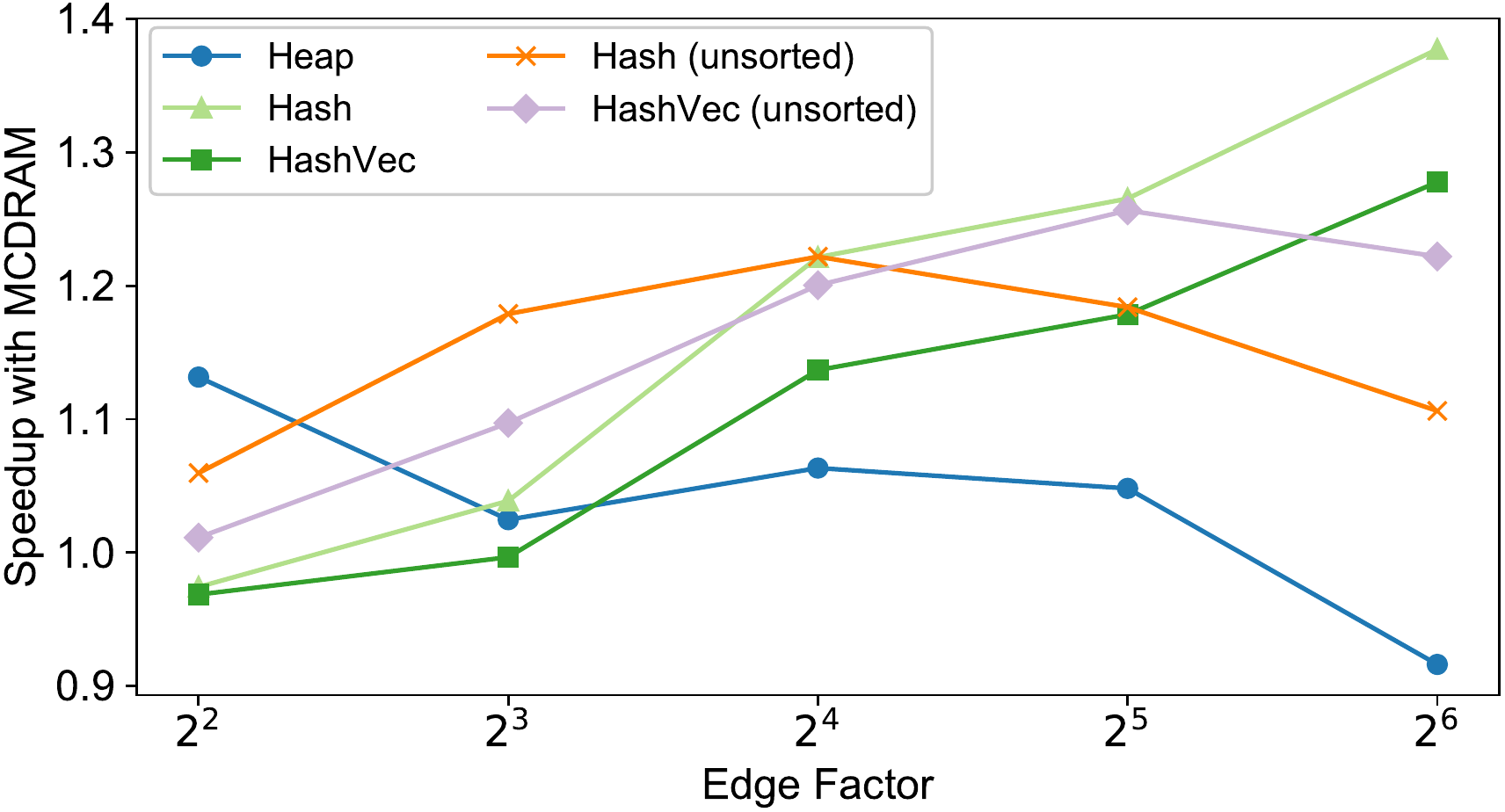}
  \caption{Speedups attained with the use of Cache mode on KNL compared to Flat mode on DDR4. G500 (scale 15) matrices are used with different edge factors.}
  \label{fig:spgemm_cache_mcdram}
 \end{center}
\end{figure}

We examine the benefit of using MCDRAM over DD4 memory by squaring G500 matrices with or without using MCDRAM on KNL.
Figure \ref{fig:spgemm_cache_mcdram} shows the speedup attained with the Cache mode against the Flat mode on DDR4 for various matrix densities.  
We observe that Hash SpGEMM algorithms can be benefitted, albeit moderately, from MCDRAM when denser matrices are multiplied.  
This observation is consistent with the benchmark shown in Figure~\ref{fig:stanza}.
The limited benefit stems from the fact that SpGEMM frequently requires indirect fine-grained memory accesses often as small as 8 bytes. 
On denser matrices, MCDRAM can still bring benefit from contiguous memory accesses of input matrices.
By contrast,  Heap SpGEMM is not benefitted from high-bandwidth  MCDRAM because of its fine-grained memory accesses.
The performance of Heap  SpGEMM even degrades when edge factor is 64 at which point the memory requirement of  Heap  SpGEMM surpasses the capacity of MCDRAM.

\subsection{Squaring a matrix}
Multiplying a sparse matrix by itself is a well-studied SpGEMM scenario. Markov clustering is an example of this case, which requires $A^2$ for a given doubly-stochastic similarity matrix. We evaluate this case extensively, under using real and synthetically generated data. For synthetic data, we provide experiments with varying density (for a fixed sized matrix) and with varying size (for a fixed density). 

\subsubsection{Scaling with Density}
\begin{figure*}[t]
 \begin{center}
    \includegraphics[width=0.95\hsize]{\fig_dir/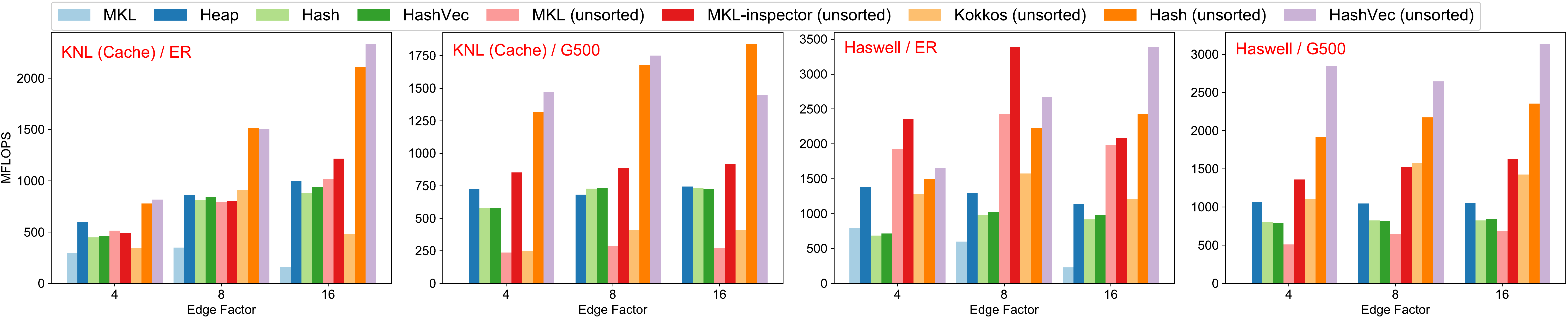}
  \caption{Scaling with increasing density (scale 16) on KNL (left) and Haswell (right)}
  \label{fig:spgemm_ef}
 \end{center}
\end{figure*}

Figure~\ref{fig:spgemm_ef} shows the result of scaling with density. 
When output is sorted, MKL's performance degrades with increasing density. 
When the output is not sorted, increased density generally translates into increased performance.
The performance of all codes except MKL increases significantly as the ER matrices get denser, but such performance gains are not so pronounced for G500 matrices. 
For G500 matrices, we see significant performance difference between KNL and Haswell results. While Hash shows superior performance on KNL, HashVector achieves much higher performance on Haswell. 
Also for G500 matrices, we see that MKL (both sorted and unsorted) and MKL-inspector achieves a peak in performance at edgecount 8, with performance degrading as the matrices get denser or sparser than that sweet spot. Hash and HashVector might not have peaked at these density ranges we experimented since they get faster as the matrices get denser. 

\subsubsection{Scaling with Input Size} 
\begin{figure*}[t]
\begin{center}
  \includegraphics[width=0.95\hsize]{\fig_dir/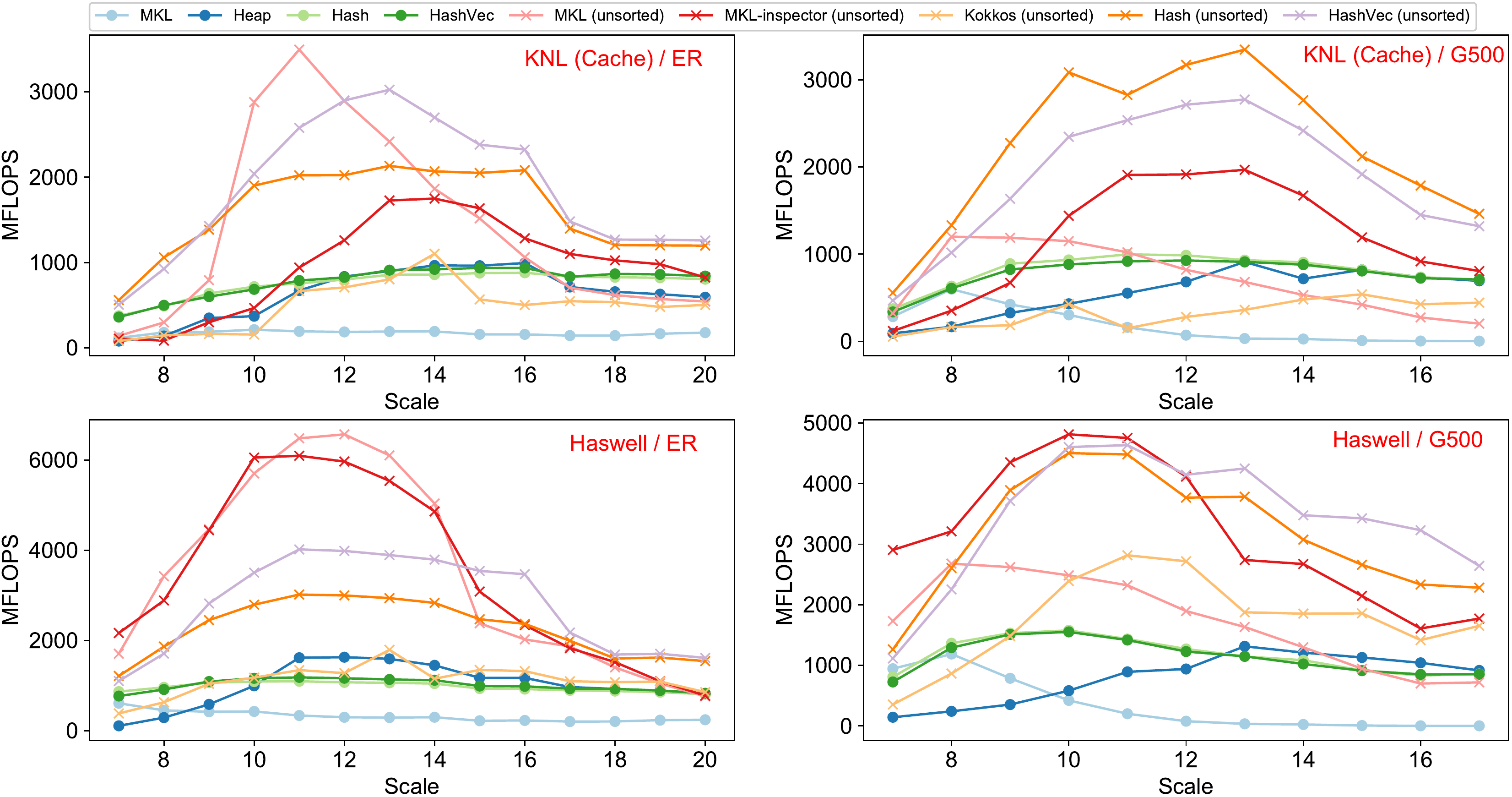}
  \caption{Scaling with size on KNL with Cache mode (top) and Haswell (bottom), both with edge factor 16}
  \label{fig:spgemm_scale}
 \end{center}
\end{figure*}

Evaluation is running on ER and G500 matrices with scaling the size from 7 to 20 or 17 respectively. The edge factor is fixed as 16. Figure~\ref{fig:spgemm_scale} shows the results on KNL (top) and Haswell (bottom).
On KNL, MKL family with unsorted output shows good performance for ER matrices with small scale. However, for large scale matrices, MKL goes down, and Heap and Hash overcome. Especially, Hash and HashVector keep high performance even for large scale matrices. This performance trend becomes more clear on Haswell. When the scale is about 13, the performance gap between sorted and unsorted is large, and it becomes smaller when the scale is getting large. This is because the cost of computation with hash table or heap becomes larger, and the advantage of removing sorting phase becomes relatively smaller.
For G500 matrices, whose non-zero elements of each row are skewed,
the performance of MKL is terrible even if its output is unsorted. Since there is no issue about load-balance in Heap and Hash kernels, they show stable performance as ER matrices.

\subsubsection{Scaling with Thread Count}
\begin{figure*}[t]
\begin{center}
  \includegraphics[width=0.9\hsize]{\fig_dir/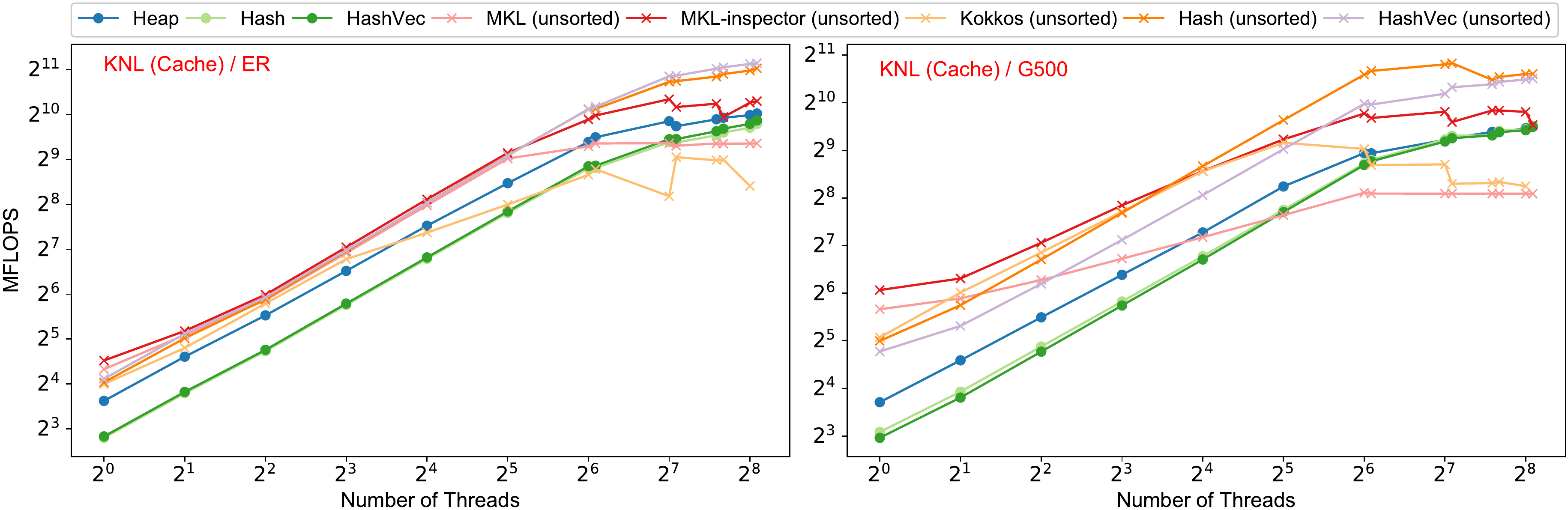}
  \caption{Strong scaling with thread count on KNL with Cache mode with ER (left) and G500 inputs (right). Data used is of scale 16 with edge factor 16}
  \label{fig:spgemm_thread}
 \end{center}
\end{figure*}

Figure~\ref{fig:spgemm_thread} shows the scalability analysis of KNL on ER and G500 matrices with scale=16 and edge factor=16. Each kernel is executed with 1, 2, 4, 8, 16, 32, 64, 68, 128, 136, 192, 204, 256 or 272 threads. We do not show the result of MKL with sorted output since it takes much longer execution time compared to other kernels. All kernels show good scalability until around 64 threads, but MKL with unsorted output has no improvement over 68 threads. On the other hand, Heap and Hash/HashVector get further improvement over 64 threads.

\subsubsection{Sensitivity to Compression Ratio on Real Matrices}
\begin{figure*}[t]
 \begin{center}
  \subfloat {
    \includegraphics[width=0.49\hsize]{\fig_dir/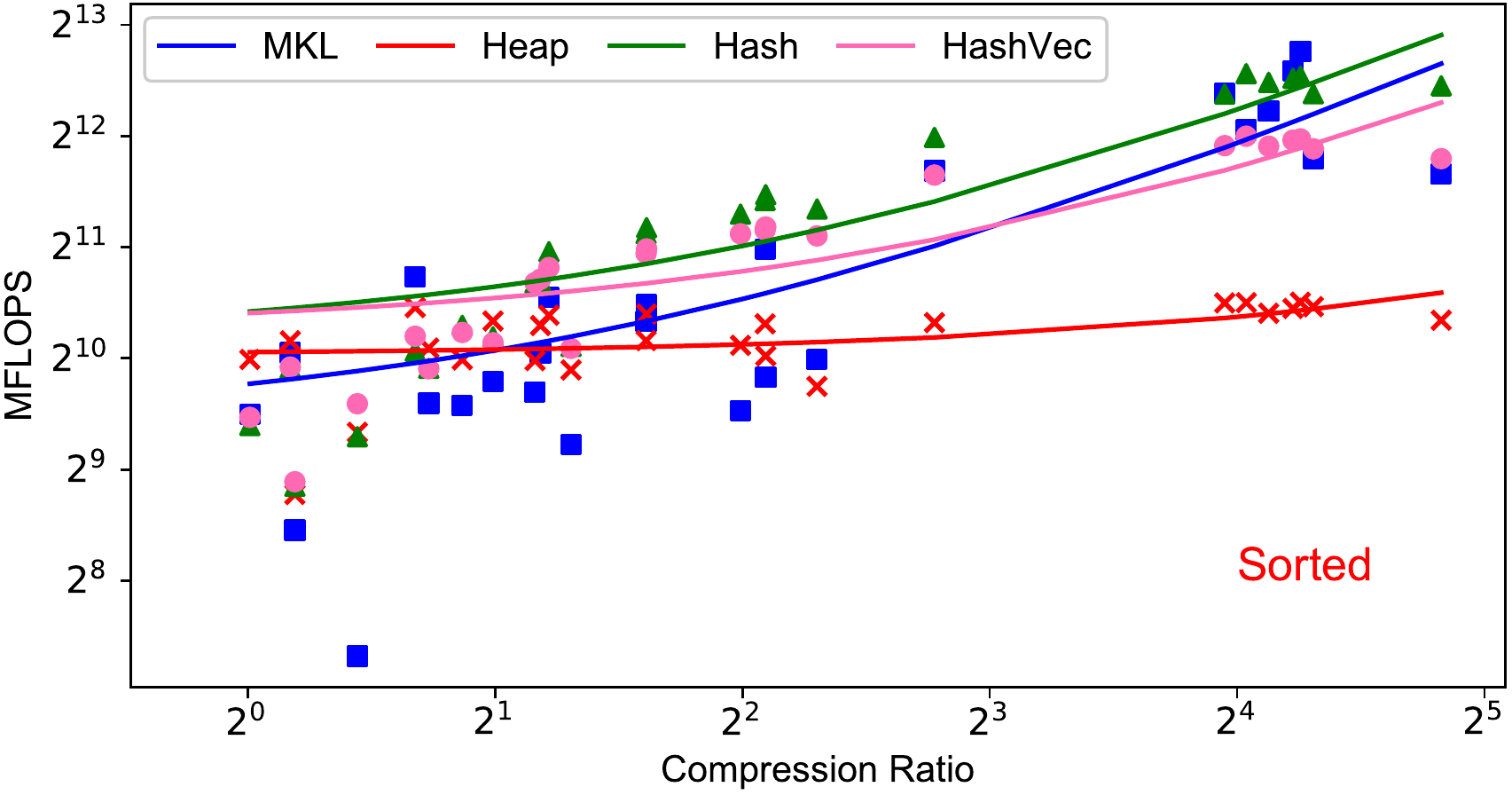}
  }
  \subfloat {
    \includegraphics[width=0.49\hsize]{\fig_dir/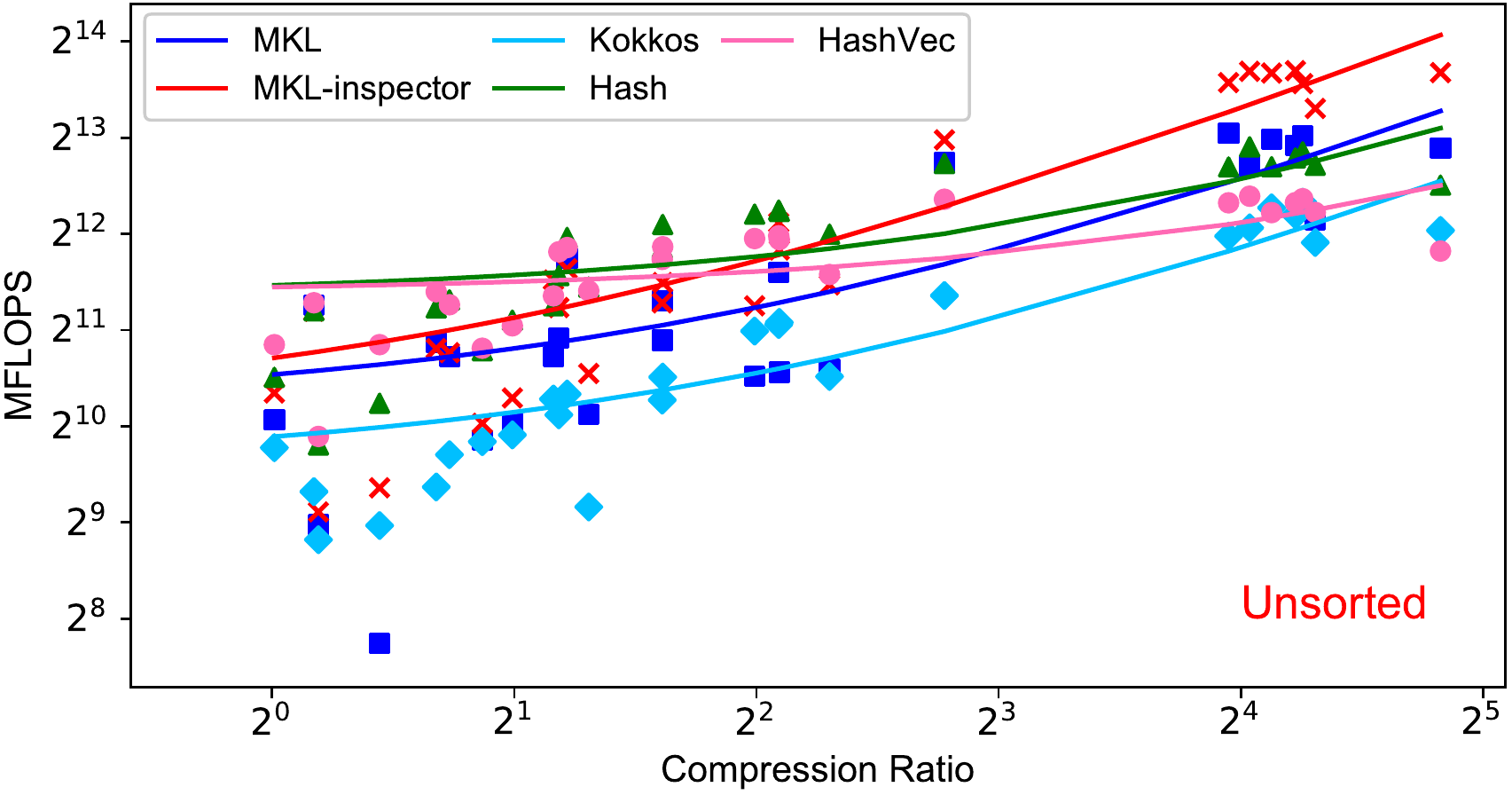}
  }
  \caption{Scaling with compression ratio of SuiteSparse matrices on KNL with Cache mode. The algorithms that operate on sorted matrices (both input \& output) are on the left and those that operate on unsorted matrices are on the right.}
  \label{fig:spgemm_florida}
 \end{center}
\end{figure*}

We evaluate SpGEMM performance on 26 real matrices listed in Table~\ref{tab:florida_mat} on KNL.
Figure~\ref{fig:spgemm_florida} shows the result with sorted output and unsorted output respectively in ascending order of compression ratio (= $\flop$ / number of non-zero elements of output). Lines in the graph are linear fitting for each kernel.
First we discuss the result with sorted matrices. The performance of Heap is stable regardless of compression ratio while MKL gets better performance with higher compression ratio. 
The matrices about graph processing with low compression ratio cause load imbalance and performance degradation on MKL.
In contrast, Hash outperforms MKL on most of matrices, and shows high performance independent from compression ratio.
For unsorted matrices, we add KokkosKernels to the evaluation, but it underperforms other kernels in this test. The performance of Hash SpGEMM is best for the matrices with low compression ratio and
MKL-inspector shows significant improvement especially for the matrices with high compression ratio.

Comparing sorted and unsorted versions of algorithm that provide the flexibility, we see consistent performance boost of keeping the output sorted. In particular, the harmonic mean of the speedups achieved operating on unsorted data over all real matrices we have studied from the SuiteSparse collection on KNL is $1.58\times$ for MKL, $1.63\times$ for Hash, and $1.68\times$ for HashVector. 

\subsubsection{Profile of Relative Performance of SpGEMM Algorithms}
\begin{figure*}[t]
 \begin{center}
  \subfloat {
    \includegraphics[width=0.4\hsize]{\fig_dir/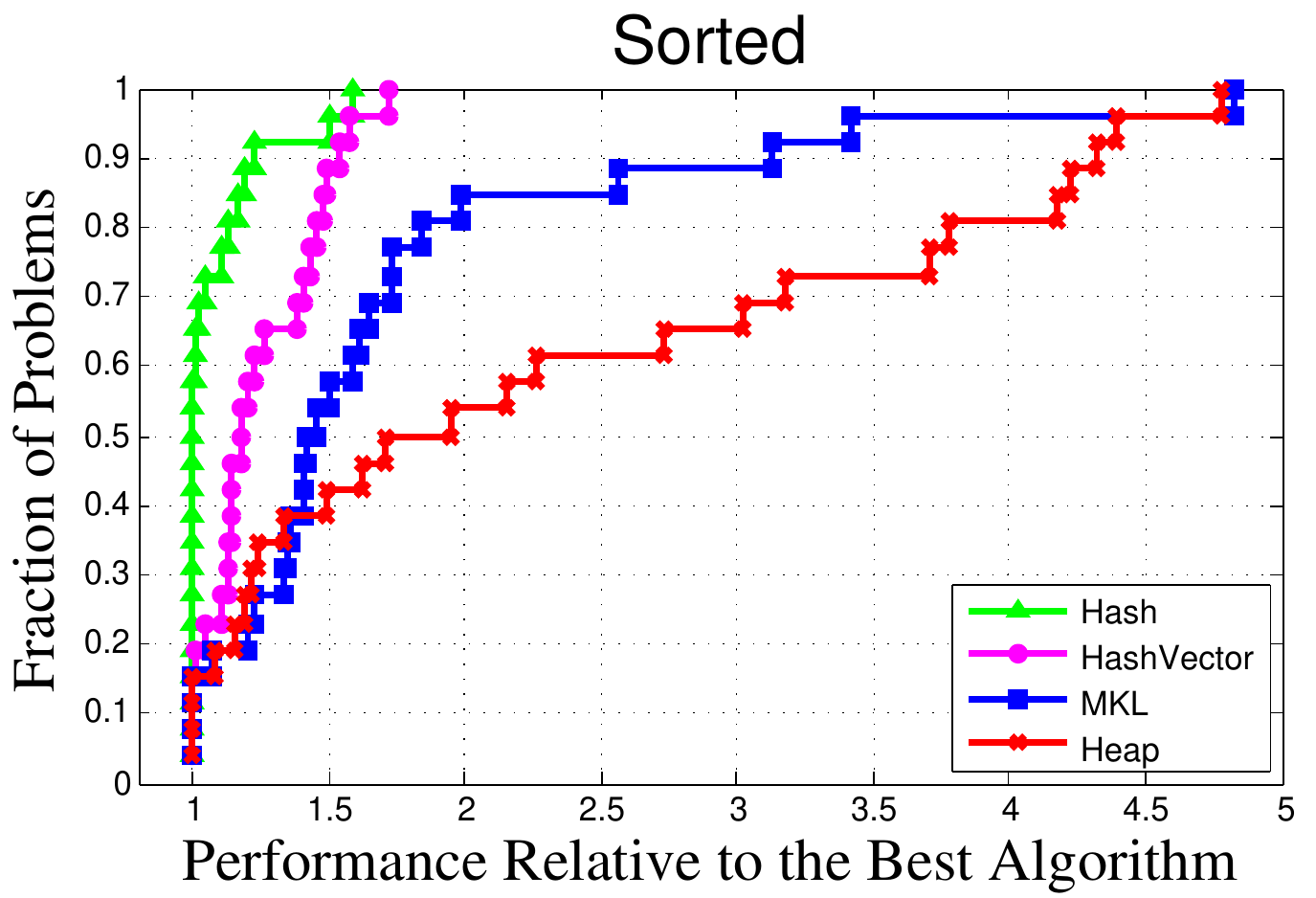}
  }
  ~~
  \subfloat {
    \includegraphics[width=0.4\hsize]{\fig_dir/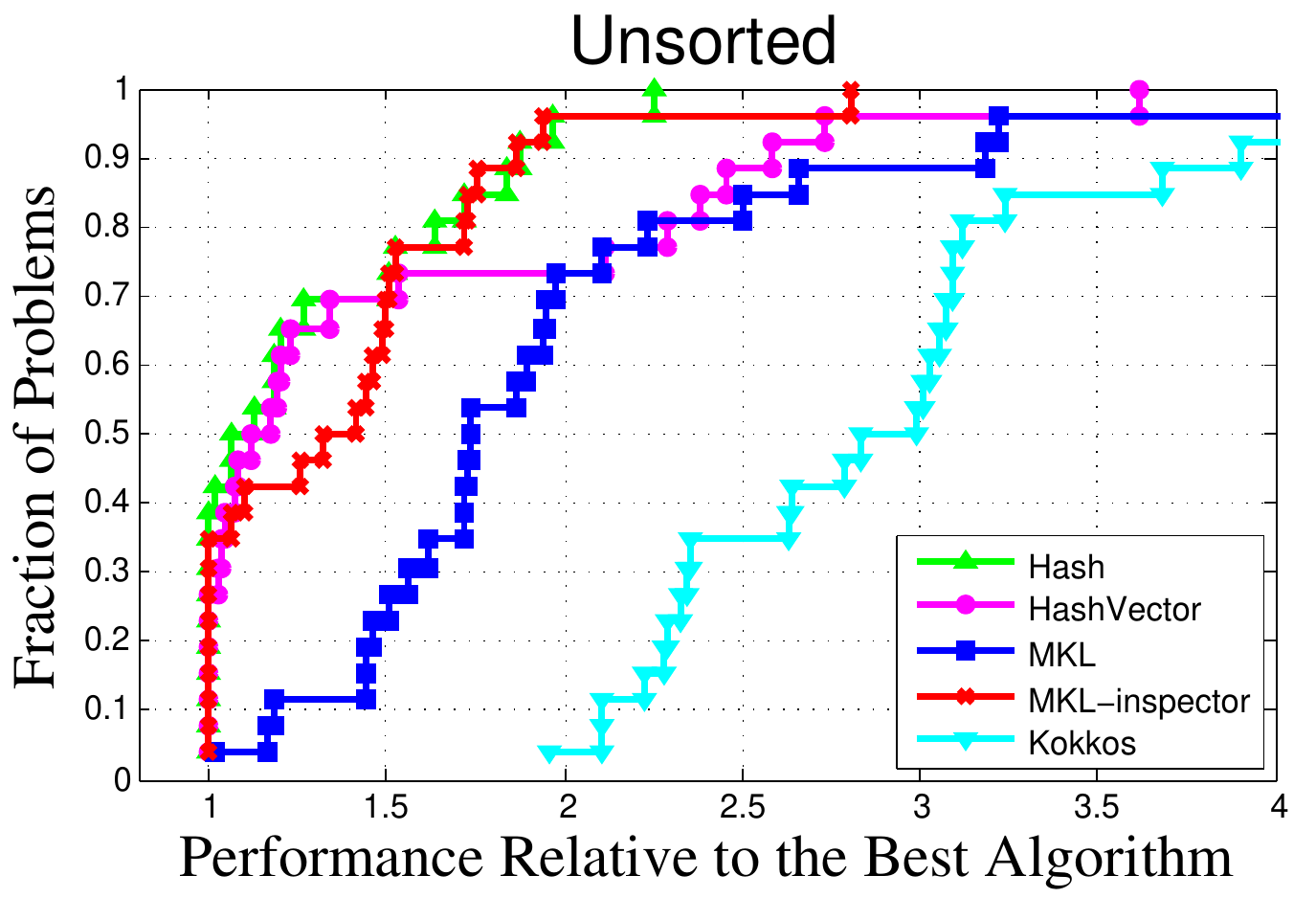}
  }
  \caption{Performance profiles of SuiteSparse matrices on KNL with Cache mode using sorted (left) and unsorted (right) algorithms.}
  \label{fig:perf_profile}
 \end{center}
\end{figure*}

We compare the relative performance of different SpGEMM algorithms with  performance profile plots~\cite{dolan2002benchmarking}. 
To profile the relative performance of algorithms,  the best performing algorithm for each problem is identified and assigned a relative score of 1.
Other algorithms are scored relative to the best performing algorithm, with a higher value denoting inferior performance for that particular problem.
For example, if algorithm A and B solve the same problem in 1 and 3 seconds, their relative performance scores will be 1 and 3, respectively.
Figure~\ref{fig:perf_profile} shows the profiles of relative performance of different SpGEMM algorithms for all 26 matrices from Table~\ref{tab:florida_mat}.
Hash is clearly the best performer for sorted matrices as it outperforms all other algorithms for 70\% matrices and  its runtime is always within $1.6\times$ of the best algorithm. 
Hash is followed by HashVector, MKL and Heap algorithms in decreasing order of overall performance. 
For unsorted matrices, Hash, HashVector and MKL-inspector all perform equally well for most matrices (each of them performs the best for about 40\% matrices).
They are followed by MKL and KokkosKernels, with the latter being the worst performer for unsorted matrices.  

\subsection{Square x Tall-skinny matrix}
\begin{figure*}[t]
 \begin{center}
  \includegraphics[width=\hsize]{\fig_dir/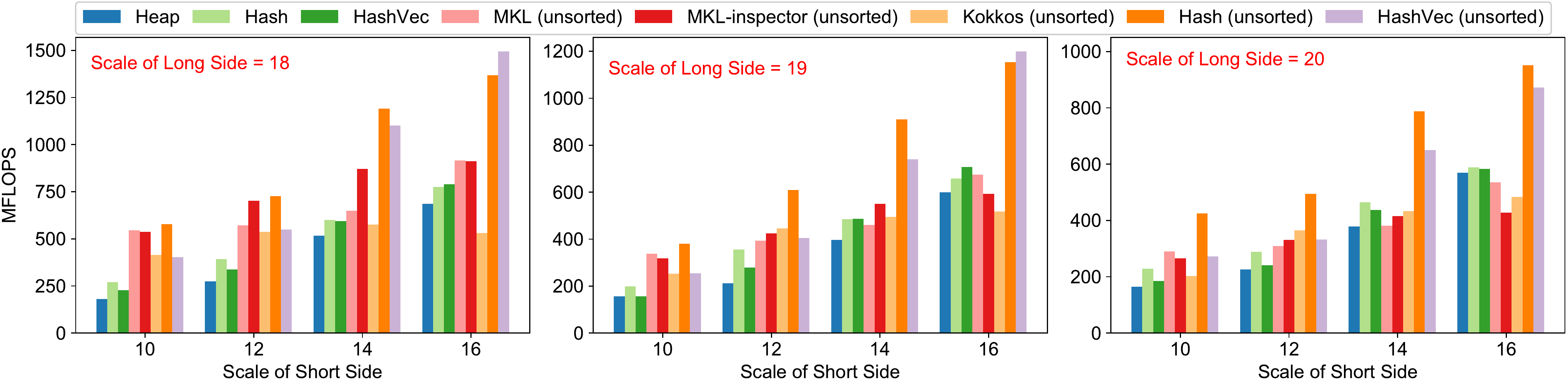}
  \caption{SpGEMM between square and tall-skinny matrices on KNL with Cache mode (scales 18, 19, and 20)}
  \label{fig:spgemm_tallskinny}
 \end{center}
\end{figure*}

Many graph processing algorithms perform multiple breadth-first searches (BFSs) in parallel, an example being Betweenness Centrality on unweighted graphs. 
In linear algebraic terms, this corresponds to multiplying a square sparse matrix with a tall-skinny one. The left-hand-side matrix represents the graph and the right-hand-side matrix represent the stack of frontiers, each column representing one BFS frontier.
In the memory-efficient implementation of the Markov clustering algorithm~\cite{hipmcl}, a matrix is multiplied with a subset of its column, representing another use case of multiplying a square matrix with a tall-skinny matrix. 
In our evaluations, we generate the tall-skinny matrix by randomly selecting columns from the graph itself. 
Figure~\ref{fig:spgemm_tallskinny} shows the result of SpGEMM between square and tall-skinny matrices.
We set scale as 18, 19 or 20 for square matrix, and as 10, 12, 14 or 16 for short size of tall-skinny matrix.
The non-zero pattern of generated matrix is G500 with edge factor=16. The result of square x tall-skinny follows that of $A^2$ (upper right in Figure~\ref{fig:spgemm_scale}). 
Both for sorted and unsorted cases, Hash or HashVec is the best performer.

\subsection{Triangle counting}
\begin{figure}[t]
 \begin{center}
  \includegraphics[width=\hsize]{\fig_dir/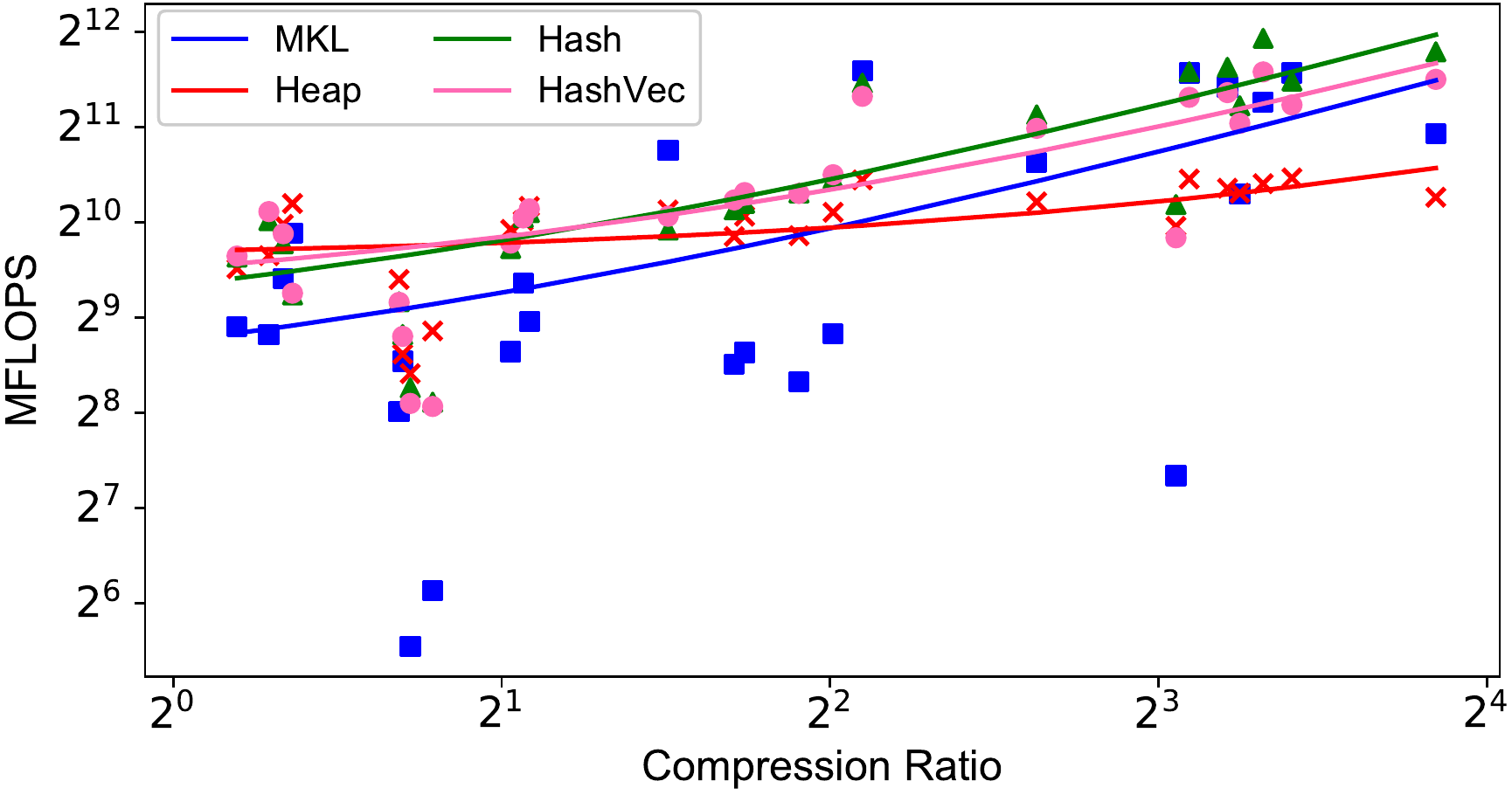}
    \caption{The performance of SpGEMM between L and U triangular matrices when used to count triangles on KNL with Cache mode}
  \label{fig:triangle_count}
 \end{center}
\end{figure}

We also evaluate the performance of SpGEMM used in triangle counting~\cite{triangles}. The original input is the adjacency matrix of an undirected graph. 
For optimal performance in triangle counting, we reorder rows with increasing number of nonzeros. The algorithm then splits the reordered matrix $A$ as $A = L + U$, where $L$ is a lower triangular matrix and $U$ is an upper triangular matrix. We evaluate the SpGEMM performance of the next step, where $L \cdot U$ is computed to generate all wedges.
After preprocessing the input matrix, we compute SpGEMM between $L$ and $U$.
Figure~\ref{fig:triangle_count} shows the result with sorted output in ascending order of compression ratio on KNL. Lines in the graph are linear fitting for each kernel.
Basically, the result shows similar performance trend to that of $A^2$. Hash and HashVector generally overwhelm MKL for any compression ratio. One big difference from $A^2$ is that Heap performs the best for inputs with low compression ratios.

\subsection{Empirical Recipe for SpGEMM on KNL}
From our evaluation results, best algorithm is different by target use case and inputs.
We summarize the recipe to choose best algorithm for SpGEMM on KNL in Table~\ref{tab:best_algo}.
The recipe for real data is based on compression ratio, which effects the dominant code. For low compression ratios, especially LxU where output is sparser, Heap shines. In the other cases, Hash and MKL-inspector dominates the high-compression ratio scenarios.
On synthetic data, our hash-table-based implementations dominate others for almost cases, and Heap works well for sparser matrices or uniform data. We note that $A^2$ for uniform input matrices shows low compression ratio.
This empirical recipe was already predicted by our analysis in Section~\ref{sec:perf_est}.

\begin{table}[]
  \centering
  \caption{Summary of best SpGEMM algorithms on KNL}
    \subfloat[Real data specified by compression ratio (CR)]{
    \small
    \begin{tabular}{ c  c | c  c }
 & & High CR ($>2$) & Low CR ($\leq 2$) \\
\hline
A x A & Sorted & Hash & Hash \\
 & Unsorted & MKL-inspector & Hash \\
\hline
L x U & Sorted & Hash & Heap \\
\hline
\hline
    \end{tabular}
    }
    \vspace{1pt}
    \subfloat[Synthetic data specified by sparsity (edge factor, EF) and non-zero pattern] {
    \small
    \begin{tabular}{ c  c | c  c  c  c}
 & & \multicolumn{2}{|c}{Sparse (EF $\leq 8$)} & \multicolumn{2}{c}{Dense (EF > 8)} \\
 & & Uniform & Skewed & Uniform & Skewed \\
\hline
A x A & Sorted & Heap & Heap & Heap & Hash\\
 & Unsorted & HashVec & HashVec & HashVec & Hash \\
\hline
TallSkinny & Sorted & - & Hash & - & HashVec \\
 & Unsorted & - & Hash & - & Hash \\
\hline
\hline
    \end{tabular}
    }
  \label{tab:best_algo}
\end{table}

\section{Conclusions}
We studied the performance of computing the multiplication of two sparse matrices on multicore and Intel KNL architectures. 
This primitive, known as SpGEMM, has recently gained attention in the GPU community, but there has been relatively less work on CPUs and other accelerators.
We have tried to fill that gap by evaluating publicly accessible implementations, including those in proprietary libraries.
From architecture point of view, we develop the optimized Heap and Hash SpGEMM algorithms for multicore and Intel KNL architectures. 
Performance evaluation shows that our optimized SpGEMM algorithms largely overcome Intel MKL and Kokkos-kernel.

Our work provides multiple recipes.
One is for the implementers of new algorithms on highly-threaded x86 architectures. We have found that the impact of memory allocation and deallocation to be significant enough to warrant optimization as without them SpGEMM performance does not scale well with increasing number of threads. We have also uncovered the impact of MCDRAM for the SpGEMM primitives. When the matrices are sparser than a threshold ($\approx 4$ nonzeros on average per row), the impact of MCDRAM is minimal because in that regime the computation becomes close to latency bound. On the other than, MCDRAM shines as matrices get denser because then SpGEMM becomes primarily bandwidth bound and can take advantage of the higher bandwidth available on MCDRAM.
The second recipe is for the users. Our results show that different codes dominate on different inputs, and we clarify which SpGEMM algorithm works well from both theoretical and empirical points of view. For example, MKL is a perfectly reasonable option for small matrices with uniform nonzero distributions. However, our heap and hash-table-based implementations dominate others for larger matrices. Similarly, compression ratio also effects the dominant code. Our results also highlight the benefits of leaving matrices (both inputs and output) unsorted whenever possible as the performance savings are significant for all codes that allow both options.
Finally, this optimization strategy for acquiring these two recipes is beneficial for optimization of SpGEMM on future architectures.

\bibliographystyle{ACM-Reference-Format}

\end{document}